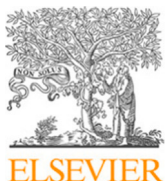

Contents lists available at ScienceDirect

# Finance Research Letters

journal homepage: www.elsevier.com/locate/frl

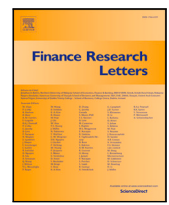

# The rise of digital finance: Financial inclusion or debt trap?

Pengpeng Yue [a], Aslihan Gizem Korkmaz [b,*], Zhichao Yin [c], Haigang Zhou [d]

[a] *Beijing Technology and Business University, Beijing, 100048, China*
[b] *Dominican University of California, San Rafael, CA, 94901-2298, USA*
[c] *Capital University of Economics and Business, Beijing, 100070, China*
[d] *Cleveland State University, Cleveland, OH, 44115-2214, USA*

A R T I C L E  I N F O

A B S T R A C T

This study focuses on the impact of digital finance on households. While digital finance has brought financial inclusion, it has also increased the risk of households falling into a debt trap. We provide evidence that supports this notion and explain the channel through which digital finance increases the likelihood of financial distress. Our results show that the widespread use of digital finance increases credit market participation. The broadened access to credit markets increases household consumption by changing the marginal propensity to consume. However, the easier access to credit markets also increases the risk of households falling into a debt trap.

## 1. Introduction

Prior literature on digital finance has mostly focused on a positive consequence of technological advancements in the finance industry, that is, financial inclusion (e.g., Banna and Alam, 2021; Beck et al., 2018; Demirgüç-Kunt et al., 2015; Hasan et al., 2021; Li et al., 2020; Lu et al., 2021; Mushtaq and Bruneau, 2019; Ren et al., 2018; Zhong and Jiang, 2021). However, the other side of the coin tells a different story where these new technologies, when inadequately regulated, may also hurt households by increasing the risk of financial distress. There are fewer studies focusing on this negative consequence of technological advancements in the finance industry. Our study aims to add to this strand of literature and examines the impact of digital finance advancements on households by focusing on both positive and negative consequences. Furthermore, we explain the channel through which digital finance may increase the likelihood of financial distress. Our goal is to provide a more complete picture on the impact of digital finance advancements on households. Our study is motivated by the need to promote digital financial literacy and prudent financial inclusion and to regulate digital financial markets.

Digital finance has a huge potential to serve people excluded from the traditional financial system (Bourreau and Valletti, 2015; Yang et al., 2020) by reducing the information asymmetry between lenders and borrowers and decreasing transaction costs (Banna and Alam, 2021; Huang and Wang, 2017; Li et al., 2020; Mishkin and Strahan, 1999; Ren et al., 2018; Xu, 2017) through the use of big data and cloud computing (Huang and Wang, 2017; Leong et al., 2017; Lu et al., 2021; Mishkin and Strahan, 1999; Xu, 2017; Yang et al., 2020) and achieving economies of scale. However, as digital finance evolves, populations with lower financial literacy gain access to complex financial products and services, some of which carry risks that may not yet be known to the market (Leong et al., 2017). Recently, the substantial increase in household debt has become a threat to the world economy and drawn attention from policymakers and researchers (Feng et al., 2019). Chen et al. (2020) posit that significant new risk might spill over from digital finance to traditional finance and that these spillovers present great challenges for policymakers.

Gabor and Brooks (2017) emphasize the role of regulation and financial literacy programs to decrease the risk of households falling into financial distress. Panos and Wilson (2020) assert that digital finance platforms such as mobile apps may damage financial well-being by triggering impulsive consumer behavior. Ozili (2018, 2020), and Xu (2017) also discuss the systemic risks imposed by digital finance on the overall finance industry. Nevertheless, these studies are qualitative in nature. Our study extends this line of literature by providing empirical evidence from China on the impact of digital finance.

Digital credit platforms can act as complements for traditional bank credit and meet the credit demand that would otherwise be unmet by banks (Hodula, 2021). Digital finance alleviates the credit constraints and increases the credit supply to Chinese SMEs (Lu et al., 2021; Sheng, 2021) and households (Yang et al., 2020; Zhong and Jiang, 2021). Thus, a higher level of financial inclusion can be achieved by digital finance. This brings us to our first hypothesis that is;

H1: Widespread use of digital finance is related to easier access to credit markets.

It has been argued in the literature that access to credit markets can increase household consumption by changing the marginal propensity to consume. Gross et al. (2020) posit that households exhibit a high marginal propensity to consume (MPC) out of transitory income shocks. Flavin (1984) shows that liquidity constraints are an important part of the observed excess sensitivity of consumption to current income, and Li et al. (2020) find digital-based financial inclusion to increase household consumption. Levchenko (2005) explains that in developing countries, consumption volatility may increase with financial liberalization. Taking easier access to the credit market as a liquidity shock, we set our second hypothesis as follows;

H2: Easier access to credit markets is related to a higher level of household consumption.

Leong et al. (2017) and Davis and Murphy (2016) discuss risks related to the lack of financial literacy by market participants. Such people may not be informed about the risks associated with using digital finance (Ozili, 2020). Feng et al. (2019) explain that financially illiterate household heads who are unaware of the consequences of their choices on debt are financially vulnerable. Analyzing the online borrowing behavior of Chinese college students, Liu and Zhang (2021) explain that due to the easy availability of online consumer credit, inflation, and the declining economy, the risks of serious financial problems are increasing. Finally, this brings us to our third hypothesis that is;

H3: Increased consumption due to the widespread use of digital finance is related to a higher possibility of falling into a debt trap.

Our results show that broadened access to digital finance increases credit market participation. The easier access to credit markets via digital platforms ushers consumption by changing the marginal propensity to consume. However, easier access to credit markets also increases the risk of financial distress.

Xu (2017) posits that due to data scarcity, the literature on digital finance is limited. Mushtaq and Bruneau (2019) assert that studies examining the impact of financial sector development with respect to developing economies are sparse and that designing a regulatory framework and consumer protection mechanism for the changing landscape of financial markets is crucial. Our contributions to the literature are threefold. First, using household data, we provide evidence at a micro-level. Second, focusing on China, we provide evidence from a country that is unique in multiple ways. China has built a comprehensive financial system (Huang and Wang, 2017), and it is the leading country in terms of the volume of digital finance worldwide (Farag and Johan, 2021; Gong et al., 2020). There is financial underservicing, especially with respect to SMEs and low-income households, and digital finance fills an important gap in the financial market in China (Huang and Wang, 2017). However, Chinese household heads exhibit lower financial literacy than those from developed countries (Feng et al., 2019). Furthermore, despite the significant progress of digital finance in China, it suffers from legal gaps and inconsistencies (Yang et al., 2020). Finally, we provide a more complete picture on the impact of digital finance by documenting both positive and negative consequences and by describing the channel through which digital finance may increase the likelihood of financial distress.

## 2. Research methods

### 2.1. Data and variables

#### 2.1.1. Data

We use the last four waves of the China Household Finance Survey (CHFS) in 2013, 2015, 2017, and 2019, and the city-level DFI index and other subindices for years 2013, 2015, and 2017 from Peking University's Digital Financial Inclusion Index (DFI) of China.

#### 2.1.2. Variables

To measure financial development, we use the total digital financial inclusion index (Total DFI index). Additionally, following Li et al. (2020) we use the two subdimensions of the Total DFI index, which measure the accessibility of internet financial services (Index of Coverage Breadth), and actual use of internet financial services (Index of Use Depth), and three subclassifications of the depth of digital finance usage which measure insurance services (Index of Insurance), investment services (Index of Investment), and credit investigation services (Index of Credit Investigation).

Based on prior literature, variables used at household level include age (Allen et al., 2016; Fungáčová and Weill, 2015; Li et al., 2020; Meyll and Pauls, 2019; Mian and Sufi, 2011), age squared (Allen et al., 2016; Fungáčová and Weill, 2015; Li et al., 2020; Meyll and Pauls, 2019), education (Allen et al., 2016; Fungáčová and Weill, 2015; Li et al., 2020; Meyll and Pauls, 2019; Ren et al.,

**Table 1**

Summary statistics. This table presents summary statistics. *Total DFI Index* is the overall digital financial inclusion index of China. *Index of coverage breadth* is a subdimension of the total DFI index, which measures the accessibility of internet financial services. *Index of use depth* is a subdimension of the total DFI index, which measures the actual use of internet financial services. *Index of insurance* is a subclassification of the depth of digital finance usage, which measures insurance services. *Index of investment* is a subclassification of the depth of digital finance usage, which measures investment services. *Index of credit investigation* is a subclassification of the depth of digital finance usage, which measures credit investigation services. *Debt Dummy* is used to identify households with outstanding loan balances. *Debt Trap* is a dummy variable equal to 1 if the survey respondent self-reports payment difficulty. *Age* is the head of household's (HOH) age. *Education* is the number of years of schooling received by the HOH. *Male* is a dummy variable equal to 1 if the HOH is male and 0 if the HOH is female. *Employed* is a dummy variable equal to 1 if the HOH is employed. *Income* is the total household income. *Consumption* is total household consumption. *Debt* is the total outstanding loan balance for each household. *Wealth* is household assets minus debt. *Rural* is a dummy variable equal to 1 if the household lives in an area categorized as rural.

Panel A: DFI

| | 2013–2017 | | | 2013 | | | 2015 | | | 2017 | | |
|---|---|---|---|---|---|---|---|---|---|---|---|---|
| | Obs | Mean | Std. | Obs | Mean | Std. | Obs | Mean | Std. | Obs | Mean | Std. |
| Total DFI index | 80,107 | 191.952 | 44.637 | 19,061 | 135.983 | 22.362 | 29,521 | 183.736 | 23.469 | 31,525 | 233.486 | 23.719 |
| Index of coverage breadth | 80,107 | 181.267 | 46.839 | 19,061 | 125.031 | 29.818 | 29,521 | 182.071 | 33.207 | 31,525 | 214.516 | 31.408 |
| Index of use depth | 80,107 | 189.140 | 57.616 | 19,061 | 137.561 | 27.109 | 29,521 | 154.828 | 25.243 | 31,525 | 252.458 | 25.965 |
| Index of insurance | 80,107 | 351.278 | 94.832 | 19,061 | 302.562 | 55.535 | 29,521 | 275.332 | 43.289 | 31,525 | 451.851 | 47.765 |
| Index of investment | 61,046 | 180.882 | 46.806 | | | | 29,521 | 139.748 | 25.285 | 31,525 | 219.400 | 23.992 |
| Index of credit investigation | 61,046 | 197.548 | 140.666 | | | | 29,521 | 59.768 | 27.878 | 31,525 | 326.571 | 56.250 |

Panel B: CHFS

| | 2013–2019 | | | 2013 | | | 2015 | | | 2017 | | | 2019 | | |
|---|---|---|---|---|---|---|---|---|---|---|---|---|---|---|---|
| | Obs | Mean | Std. | Obs | Mean | Std. | Obs | Mean | Std. | Obs | Mean | Std. | Obs | Mean | Std. |
| Debt Dummy | 128,452 | 0.287 | 0.452 | 23,565 | 0.302 | 0.459 | 34,827 | 0.294 | 0.455 | 37,616 | 0.287 | 0.452 | 32,444 | 0.268 | 0.443 |
| Debt Trap | 104,887 | 0.042 | 0.201 | | | | 34,827 | 0.040 | 0.196 | 37,616 | 0.044 | 0.206 | 32,444 | 0.041 | 0.199 |
| Age | 128,452 | 54.424 | 14.332 | 23,565 | 51.555 | 14.448 | 34,827 | 53.360 | 14.381 | 37,616 | 55.359 | 14.312 | 32,444 | 56.566 | 13.758 |
| Education | 128,452 | 9.131 | 4.466 | 23,565 | 9.263 | 4.503 | 34,827 | 9.110 | 4.508 | 37,616 | 9.184 | 4.470 | 32,444 | 8.996 | 4.384 |
| Male | 128,452 | 0.765 | 0.424 | 23,565 | 0.754 | 0.430 | 34,827 | 0.755 | 0.430 | 37,616 | 0.792 | 0.406 | 32,444 | 0.751 | 0.433 |
| Employed | 128,452 | 0.641 | 0.480 | 23,565 | 0.671 | 0.470 | 34,827 | 0.646 | 0.478 | 37,616 | 0.614 | 0.487 | 32,444 | 0.645 | 0.478 |
| Income | 128,452 | 83,750 | 101,141 | 23,565 | 63,063 | 76,449 | 34,827 | 67,535 | 89,282 | 37,616 | 84,497 | 103,422 | 32,444 | 115,316 | 117,081 |
| Consumption | 128,452 | 62,206 | 67,501 | 23,565 | 48,434 | 42,958 | 34,827 | 55,457 | 56,702 | 37,616 | 57,742 | 53,309 | 32,444 | 84,631 | 96,129 |
| Debt | 128,452 | 39,705 | 136,210 | 23,565 | 29,861 | 98,706 | 34,827 | 39,075 | 147,424 | 37,616 | 43,902 | 141,523 | 32,444 | 42,667 | 140,458 |
| Wealth | 128,452 | 902,596 | 1,557,362 | 23,565 | 735,398 | 1,216,327 | 34,827 | 831,240 | 1,359,028 | 37,616 | 1,052,485 | 1,823,746 | 32,444 | 926,850 | 1,623,851 |
| Rural | 128,452 | 0.321 | 0.467 | 23,565 | 0.314 | 0.464 | 34,827 | 0.309 | 0.462 | 37,616 | 0.307 | 0.461 | 32,444 | 0.356 | 0.479 |

2018), male/female classification (Allen et al., 2016; Fungáčová and Weill, 2015; Meyll and Pauls, 2019; Mian and Sufi, 2011; Ren et al., 2018), employment (Allen et al., 2016; Jack and Suri, 2014; Ren et al., 2018), income (Allen et al., 2016; Demirgüç-Kunt and Klapper, 2013; Fungáčová and Weill, 2015; Li et al., 2020; Mian and Sufi, 2011), assets (Li et al., 2020), wealth (Li et al., 2020; Meyll and Pauls, 2019), rural/urban residence classification (Allen et al., 2016; Jack and Suri, 2014; Li et al., 2020), consumption (Jack and Suri, 2014; Li et al., 2020), existence of unpaid loans (Samaratunge et al., 2020) to proxy for access to credit markets, and self-reported financial distress (Gathergood, 2012) to proxy for debt trap.

To reduce the impact of outliers, we use winsorizing at 1% and 99% levels on continuous variables. Table 1 reports summary statistics for variables used in our analyses.

### 2.2. Models

To test our first hypothesis, we use Eq. (1):

$$Debt\,Dummy_{it} = \alpha + \beta ln(DFI)_{it} + X_{it}\gamma + c_i + \mu_{it} \qquad (1)$$

Where $Debt\,Dummy_{it}$ is a dummy variable equal to 1 if the household has an outstanding loan balance, $\alpha$ is the intercept, $ln(DFI)_{it}$ is the natural logarithm of the DFI index or a subindex, $X_{it}$ is the vector of control variables, $c_i$ represents the confounding variables, and $\mu_{it}$ is the error term.

To test our second hypothesis, we use Eq. (2):

$$ln(Consumption)_{it} = \alpha + \beta_1 ln(DFI)_{it} + \beta_2 ln(DFI)_{it} \times ln(Income)_{it} + \beta_3 ln(Income)_{it} + X_{it}\gamma + c_i + \mu_{it} \qquad (2)$$

Where $ln(Consumption)_{it}$ is the natural logarithm of total household consumption, $\alpha$ is the intercept, and $ln(DFI)_{it}$ is the natural logarithm of the DFI index.

We also analyze the relationship between debt and consumption using Eq. (3):

$$ln(Consumption)_{it} = \alpha + \beta_1 Debt\,Dummy_{it} + \beta_2 Debt\,Dummy_{it} \times ln(Income)_{it} + \beta_3 ln(Income)_{it} + X_{it}\gamma + c_i + \mu_{it} \qquad (3)$$

**Table 2**
Digital Finance and Credit Market Participation. This table reports the results of our tests analyzing the impact of digital finance on financial inclusion.

| | (1) FE | (2) FE | (3) FE | (4) FE | (5) FE | (6) FE |
|---|---|---|---|---|---|---|
| ln(Total DFI index) | 0.0293*** | | | | | |
| | (0.0093) | | | | | |
| ln(Index of coverage breadth) | | 0.0288*** | | | | |
| | | (0.0091) | | | | |
| ln(Index of use depth) | | | 0.0187*** | | | |
| | | | (0.0069) | | | |
| ln(Index of insurance) | | | | 0.0186** | | |
| | | | | (0.0074) | | |
| ln(Index of investment) | | | | | 0.0141* | |
| | | | | | (0.0085) | |
| ln(Index of credit investigation) | | | | | | 0.0045** |
| | | | | | | (0.0020) |
| Age | −0.0018 | −0.0018 | −0.0017 | −0.0016 | 0.0022 | 0.0022 |
| | (0.0020) | (0.0020) | (0.0020) | (0.0020) | (0.0024) | (0.0024) |
| Age squared | −0.0011 | −0.0010 | −0.0011 | −0.0011 | −0.0040* | −0.0040* |
| | (0.0018) | (0.0018) | (0.0018) | (0.0018) | (0.0022) | (0.0022) |
| Education | −0.0005 | −0.0004 | −0.0003 | −0.0002 | 0.0005 | 0.0004 |
| | (0.0013) | (0.0013) | (0.0013) | (0.0013) | (0.0014) | (0.0014) |
| Male | −0.0159* | −0.0156* | −0.0163* | −0.0163* | −0.0166* | −0.0168* |
| | (0.0092) | (0.0092) | (0.0093) | (0.0093) | (0.0100) | (0.0100) |
| Employed | 0.0081 | 0.0082 | 0.0080 | 0.0078 | 0.0034 | 0.0036 |
| | (0.0065) | (0.0065) | (0.0065) | (0.0065) | (0.0081) | (0.0081) |
| ln(Income) | 0.0075*** | 0.0076*** | 0.0073*** | 0.0073*** | 0.0084*** | 0.0082*** |
| | (0.0012) | (0.0012) | (0.0012) | (0.0012) | (0.0016) | (0.0016) |
| ln(Wealth) | 0.0086*** | 0.0085*** | 0.0089*** | 0.0090*** | 0.0092*** | 0.0092*** |
| | (0.0021) | (0.0021) | (0.0021) | (0.0021) | (0.0025) | (0.0025) |
| Rural | 0.0332 | 0.0334 | 0.0332 | 0.0327 | 0.0214 | 0.0216 |
| | (0.0269) | (0.0269) | (0.0269) | (0.0269) | (0.0294) | (0.0294) |
| Year | Yes | Yes | Yes | Yes | Yes | Yes |
| Observations | 80,107 | 80,107 | 80,107 | 80,107 | 61,046 | 61,046 |
| $R^2$ | 0.0043 | 0.0043 | 0.0042 | 0.0042 | 0.0045 | 0.0047 |
| Adjusted $R^2$ | 0.0042 | 0.0042 | 0.0041 | 0.0041 | 0.0044 | 0.0045 |
| F-test | 14.90 | 14.92 | 14.72 | 14.69 | 9.34 | 9.54 |

*Indicates significance at the 10% level.
**Indicates significance at the 5% level.
***Indicates significance at the 1% level.

Next, we test our third hypothesis using Eq. (4):

$$DebtTrap_{it} = \alpha + \beta ln(DFI)_{it} + X_{it}\gamma + c_i + \mu_{it} \quad (4)$$

Where $DebtTrap_{it}$ is a dummy variable equal to 1 if the household has financial distress.

## 3. Results

### 3.1. Digital finance and credit market participation

Using Eq. (1), we investigate whether digital finance increases financial inclusion by increasing the likelihood of getting a loan. Table 2 shows the results of our analysis. A 1% increase in the DFI index is related to a 2.93% increase in the likelihood of households getting a loan and a 10.21% increase in the mean value of the debt dummy variable. In unreported tests, we find that a 1% increase in the DFI index causes a 17,959 RMB increase in the debt level. These results suggest that digital finance makes it easier for households to gain access to credit markets.

### 3.2. Credit market participation and household consumption behavior

Next, we examine the impact of broadened access to credit markets on household consumption behavior by using Eq. (2) in Columns (1) and (2), and Eq. (3) in Columns (3) and (4) of Table 3. A 1% increase in the DFI index is related to a 27.29% increase in household consumption and a 4.30% increase in the marginal propensity to consume. In unreported tests, we further divide the Total DFI index into three categories as high, medium, and low. For low, medium, and high values of the Total DFI index, when income increases by 1%, consumption increases by 3.35%, 4.78%, and 5.69%, respectively. These results indicate that the higher the total DFI index, the stronger the impact of income on consumption.

Then, using Eq. (3), we try to see if debt changes the marginal propensity to consume. However, in Column (4), the insignificant coefficient on the interaction term shows that debt does not have a significant impact.

**Table 3**
Credit Market Participation and Household Consumption Behavior. This table reports the results of our tests analyzing the impact of digital finance and debt on household consumption behavior.

|  | (1) FE | (2) FE | (3) FE | (4) FE |
| --- | --- | --- | --- | --- |
| ln(Total DFI Index) | 0.2729*** | −0.1674** |  |  |
|  | (0.0132) | (0.0758) |  |  |
| ln(Total DFI Index)* ln(Income) |  | 0.0430*** |  |  |
|  |  | (0.0072) |  |  |
| Debt Dummy |  |  | 0.1212*** | 0.1163*** |
|  |  |  | (0.0070) | (0.0372) |
| Debt Dummy* ln(Income) |  |  |  | 0.0005 |
|  |  |  |  | (0.0035) |
| Age | 0.0103*** | 0.0095*** | 0.0172*** | 0.0172*** |
|  | (0.0031) | (0.0031 | (0.0024) | (0.0024) ) |
| Square of Age | −0.0144*** | −0.0137*** | −0.0142*** | −0.0142*** |
|  | (0.0029) | (0.0029) | (0.0022) | (0.0022) |
| Education | 0.0069*** | 0.0067*** | 0.0200*** | 0.0200*** |
|  | (0.0020) | (0.0020) | (0.0015) | (0.0015) |
| Male | 0.0763*** | 0.0732*** | 0.0191* | 0.0191* |
|  | (0.0131) | (0.0131) | (0.0099) | (0.0099) |
| Employed | −0.0240** | −0.0244** | −0.0216*** | −0.0216*** |
|  | (0.0099) | (0.0099) | (0.0081) | (0.0081) |
| ln(Income) | 0.0431*** | −0.1789*** | 0.0720*** | 0.0719*** |
|  | (0.0019) | (0.0373) | (0.0018) | (0.0019) |
| ln(Wealth) | 0.0854*** | 0.0840*** | 0.0854*** | 0.0854*** |
|  | (0.0033) | (0.0033) | (0.0026) | (0.0026) |
| Rural | −0.1313*** | −0.1301*** | −0.0228** | −0.0228** |
|  | (0.0427) | (0.0427) | (0.0091) | (0.0091) |
| Year | Yes | Yes | Yes | Yes |
| Observations | 80,107 | 80,107 | 128,577 | 128,577 |
| R−squared | 0.0818 | 0.0828 | 0.0822 | 0.0822 |
| Adjusted $R^2$ | 0.0817 | 0.0827 | 0.0822 | 0.0822 |
| F-test | 253.68 | 234.68 | 439.49 | 383.69 |
| Lincom test |  | −0.2067** |  | 0.1158*** |
|  |  | (0.0130) |  | (0.0040) |

*Indicates significance at the 10% level.
**Indicates significance at the 5% level.
***Indicates significance at the 1% level.

### 3.3. Digital finance and household financial distress

In Table 4, we analyze the possibility of falling into financial distress by using Eq. (4). A 1% increase in the DFI index is related to a 2.90% increase in the likelihood of households falling into a debt trap and a 69% increase in the mean value of the debt trap variable.

### 3.4. Endogeneity

One may argue that households who are more likely to participate in financial markets are also more likely to be involved in digital finance. Li et al. (2020) discuss the endogeneity problem that arises due to this reverse causality issue and suggest using the number of mobile phones as an instrument for digital finance variable. Thus, in this section, using the ratio of households who own smartphones in a city as an instrumental variable, we reinvestigate the relationship modeled in Eq. (4). The positive significant coefficients of the DFI indices are in line with our results in Section 3.3, meaning that our results are not driven by endogeneity bias (see Table 5).

### 4. Conclusion

In this study, we investigate the impact of digital finance on households. Our results show that the wider use of digital finance increases credit market participation. Easier access to credit markets increases the marginal propensity to consume out of liquidity and stimulates consumption. However, increased borrowing also increases the risk of financial distress.

Research shows that the financial literacy of Chinese households is low, and financially illiterate households are unaware of the consequences of their choices on debt (Feng et al., 2019). Thus, the first policy implication of our results is that policymakers should provide people with appropriate digital financial literacy (Banna and Alam, 2021). However, Gathergood (2012) finds lack of self-control in addition to financial illiteracy to be positively associated with non-payment of consumer credit and self-reported excessive financial burdens of debt. The study explains that although financial literacy might be improved through financial education, individuals cannot be educated on self-control. Thus, the second policy implication of our results is that the policymakers need to restrict the credit available by controlling for loan purposes (Meyll and Pauls, 2019).

Table 4

Digital Finance and Household Financial Distress. This table reports the results of our tests analyzing the risk of falling into a debt trap.

|  | (1) FE | (2) FE | (3) FE | (4) FE | (5) FE | (6) FE |
|---|---|---|---|---|---|---|
| ln(Total DFI index ) | 0.0290*** <br> (0.0089) |  |  |  |  |  |
| ln(Index of coverage breadth ) |  | 0.0398*** <br> (0.0126) |  |  |  |  |
| ln(Index of use depth) |  |  | 0.0159*** <br> (0.0044) |  |  |  |
| ln(Index of insurance ) |  |  |  | 0.0147*** <br> (0.0042) |  |  |
| ln(Index of investment) |  |  |  |  | 0.0162*** <br> (0.0047) |  |
| ln(Index of credit investigation ) |  |  |  |  |  | 0.0042*** <br> (0.0011) |
| Age | 0.0022* <br> (0.0012) | 0.0022* <br> (0.0012) | 0.0021* <br> (0.0012) | 0.0021* <br> (0.0012) | 0.0021* <br> (0.0012) | 0.0021* <br> (0.0012) |
| Age squared | −0.0022** <br> (0.0011) | −0.0022** <br> (0.0011) | −0.0022** <br> (0.0011) | −0.0022** <br> (0.0011) | −0.0022** <br> (0.0011) | −0.0022** <br> (0.0011) |
| Education | 0.0002 <br> (0.0007) | 0.0002 <br> (0.0007) | 0.0002 <br> (0.0007) | 0.0002 <br> (0.0007) | 0.0002 <br> (0.0007) | 0.0002 <br> (0.0007) |
| Male | −0.0023 <br> (0.0049) | −0.0023 <br> (0.0049) | −0.0025 <br> (0.0049) | −0.0024 <br> (0.0049) | −0.0024 <br> (0.0049) | −0.0024 <br> (0.0049) |
| Employed | −0.0043 <br> (0.0045) | −0.0043 <br> (0.0045) | −0.0043 <br> (0.0045) | −0.0043 <br> (0.0045) | −0.0043 <br> (0.0045) | −0.0042 <br> (0.0045) |
| ln(Income) | −0.0001 <br> (0.0009) | −0.0001 <br> (0.0009) | −0.0002 <br> (0.0009) | −0.0001 <br> (0.0009) | −0.0001 <br> (0.0009) | −0.0002 <br> (0.0009) |
| ln(Wealth) | −0.0014 <br> (0.0014) | −0.0014 <br> (0.0014) | −0.0013 <br> (0.0014) | −0.0014 <br> (0.0014) | −0.0013 <br> (0.0014) | −0.0014 <br> (0.0014) |
| Rural | 0.0179 <br> (0.0148) | 0.0182 <br> (0.0148) | 0.0177 <br> (0.0148) | 0.0174 <br> (0.0148) | 0.0176 <br> (0.0148) | 0.0177 <br> (0.0148) |
| Year | Yes | Yes | Yes | Yes | Yes | Yes |
| Observations | 61,046 | 61,046 | 61,046 | 61,046 | 61,046 | 61,046 |
| $R^2$ | 0.0010 | 0.0010 | 0.0012 | 0.0011 | 0.0011 | 0.0013 |
| Adjusted $R^2$ | 0.0009 | 0.0009 | 0.0011 | 0.0010 | 0.0010 | 0.0012 |
| F-test | 2.13 | 2.04 | 2.41 | 2.29 | 2.27 | 2.46 |

\*Indicates significance at the 10% level.
\*\*Indicates significance at the 5% level.
\*\*\*Indicates significance at the 1% level.

Most small lending platforms in China prioritize expanding their businesses, operate without applying appropriate financial techniques, and in the end, they have to bear risks they do not know how to manage (Xu, 2017). This creates a significant risk that might spill over from digital finance to traditional finance (Chen et al., 2020). Accordingly, the third policy implication of our results is that the policymakers should improve customer protections (Liao et al., 2020), ensure market transparency, competition, and fair pricing (Yang et al., 2018) to reduce the risks associated with digital finance system (Ozili, 2020).

**CRediT authorship contribution statement**

**Pengpeng Yue:** Conceptualization, Methodology, Software, Formal analysis, Data curation, Writing – original draft. **Aslihan Gizem Korkmaz:** Conceptualization, Methodology, Writing – original draft, Writing – review & editing. **Zhichao Yin:** Supervision, Resources. **Haigang Zhou:** Supervision, Methodology.

**Declaration of competing interest**

The authors declare that they have no known competing financial interests or personal relationships that could have appeared to influence the work reported in this paper.

**Compliance with ethical standards**

**Funding**

This work has been financially supported by the National Natural Science Foundation of China (72103010)

**Table 5**
Robustness Test: Instrumental variable. This table reports the results of our robustness test using smartphone ownership as the instrumental variable.

|  | (1) FE 2SLS | (2) FE 2SLS | (3) FE 2SLS | (4) FE 2SLS | (5) FE 2SLS | (6) FE 2SLS |
| --- | --- | --- | --- | --- | --- | --- |
| ln(Total DFI index) | 0.0311*** |  |  |  |  |  |
|  | (0.0118) |  |  |  |  |  |
| ln(Index of coverage breadth) |  | 0.0442*** |  |  |  |  |
|  |  | (0.0168) |  |  |  |  |
| ln(Index of use depth) |  |  | 0.0151*** |  |  |  |
|  |  |  | (0.0057) |  |  |  |
| ln(Index of insurance) |  |  |  | 0.0152*** |  |  |
|  |  |  |  | (0.0058) |  |  |
| ln(Index of investment) |  |  |  |  | 0.0162*** |  |
|  |  |  |  |  | (0.0061) |  |
| ln(Index of credit investigation) |  |  |  |  |  | 0.0040*** |
|  |  |  |  |  |  | (0.0015) |
| Age | 0.0021* | 0.0021* | 0.0021* | 0.0021* | 0.0021* | 0.0021* |
|  | (0.0012) | (0.0012) | (0.0012) | (0.0012) | (0.0012) | (0.0012) |
| Age squared | −0.0022** | −0.0022** | −0.0022** | −0.0022** | −0.0022** | −0.0022** |
|  | (0.0011) | (0.0011) | (0.0011) | (0.0011) | (0.0011) | (0.0011) |
| Education | 0.0002 | 0.0002 | 0.0002 | 0.0002 | 0.0002 | 0.0002 |
|  | (0.0007) | (0.0007) | (0.0007) | (0.0007) | (0.0007) | (0.0007) |
| Male | −0.0024 | −0.0024 | −0.0024 | −0.0025 | −0.0024 | −0.0023 |
|  | (0.0052) | (0.0052) | (0.0052) | (0.0052) | (0.0052) | (0.0052) |
| Employed | −0.0043 | −0.0043 | −0.0043 | −0.0043 | −0.0043 | −0.0043 |
|  | (0.0041) | (0.0041) | (0.0041) | (0.0041) | (0.0041) | (0.0041) |
| ln(Income) | −0.0001 | −0.0001 | −0.0001 | −0.0001 | −0.0001 | −0.0001 |
|  | (0.0008) | (0.0008) | (0.0008) | (0.0008) | (0.0008) | (0.0008) |
| ln(Wealth) | −0.0013 | −0.0013 | −0.0014 | −0.0014 | −0.0013 | −0.0014 |
|  | (0.0013) | (0.0013) | (0.0013) | (0.0013) | (0.0013) | (0.0013) |
| Rural | 0.0179 | 0.0183 | 0.0177 | 0.0174 | 0.0176 | 0.0177 |
|  | (0.0158) | (0.0158) | (0.0158) | (0.0158) | (0.0158) | (0.0158) |
| Year | Yes | Yes | Yes | Yes | Yes | Yes |
| Observations | 61,046 | 61,046 | 61,046 | 61,046 | 61,046 | 61,046 |
| $R^2$ | 0.0114 | 0.0033 | 0.0158 | 0.0163 | 0.0146 | 0.0178 |
| F value of first stage | 8146.23 | 3815.22 | 10,241.10 | 8923.82 | 8312.99 | 9966.17 |
| Cragg-Donald Wald F | 1.0e+04 | 5956.65 | 1.1e+04 | 9950.64 | 1.0e+04 | 1.2e+04 |
| Kleibergen-Paap rk Wald F | 8146.23 | 3815.22 | 1.0e+04 | 8923.82 | 8312.99 | 9966.17 |
| Wald Chi2 | 3604.76 | 3604.69 | 3605.34 | 3605.00 | 3605.10 | 3605.69 |

*Indicates significance at the 10% level.
**Indicates significance at the 5% level.
***Indicates significance at the 1% level.